\documentclass[12pt,preprint]{emulateapj}

\def\lap{\lower.5ex\hbox{$\; \buildrel < \over \sim \;$}}
\def\gap{\lower.5ex\hbox{$\; \buildrel > \over \sim \;$}}
\def\Msun{\hbox{M$_{\odot}$}}
\def\Lsun{\hbox{L$_{\odot}$}}
\def\micron{$\mu$m}
\def\kms{km s$^{-1}$}
\def\Msolaryr{\Msun yr$^{-1}$}
\def\LIR{L$\mathrm{_{IR}}$}
\def\rtwo{R$\mathrm{_{200}}$}
\def\mtwo{M$\mathrm{_{200}}$}
\usepackage{graphicx}
\usepackage{natbib}
\begin{document}
\title{A WISE View of Star Formation in Local Galaxy Clusters}
\author{Sun Mi Chung\altaffilmark{1}, Peter R. Eisenhardt\altaffilmark{2}, Anthony H. Gonzalez\altaffilmark{1}, Spencer A. Stanford\altaffilmark{3}, Mark Brodwin\altaffilmark{4}, Daniel Stern\altaffilmark{2}, Thomas Jarrett\altaffilmark{5}}

\altaffiltext{1}{Department of Astronomy, University of Florida, Gainesville, FL 32611-2055, USA; schung@astro.ufl.edu}
\altaffiltext{2}{Jet Propulsion Laboratory, California Institute of Technology, Pasadena, CA 91109, USA}
\altaffiltext{3}{Department of Physics, University of California, One Shields Avenue, Davis, CA  95616, USA}
\altaffiltext{4}{Harvard-Smithsonian Center for Astrophysics, 60 Garden Street, Cambridge, MA 02138, USA} 
\altaffiltext{5}{Infrared Processing and Analysis Center, California Institute of Technology, Pasadena, CA 91125, USA}

\begin{abstract}

We present results from a systematic study of star formation in local galaxy clusters using 22$\mu$m data from the Wide-field Infrared Survey Explorer (WISE). The 69 systems in our sample are drawn from the Cluster Infall Regions Survey (CIRS), and all have robust mass determinations. The all-sky WISE data enables us to quantify the amount of star formation, as traced by 22$\mu$m, as a function of radius well beyond \rtwo, and investigate the dependence of total star formation rate upon cluster mass. We find that the fraction of star-forming galaxies increases with cluster radius but remains below the field value even at 3\rtwo. We also find that there is no strong correlation between the mass-normalized total specific star formation rate and cluster mass, indicating that the mass of the host cluster does not strongly influence the total star formation rate of cluster members.

\end{abstract}

\keywords{galaxies: clusters: general --- galaxies: evolution}


\section{Introduction}

It is well-established that the fraction of star-forming galaxies declines as a function of increasing local galaxy density in the low redshift universe.  Also known as the star formation-density relation \citep{lewis2002,gomez2003,balogh2004}, this correlation has been confirmed in many studies, primarily using optical and UV data to trace star formation in massive galaxy clusters and field environments.  Mid-infrared data from the Infrared Satellite Observatory (ISO) and the Multi- band Imaging Photometer for Spitzer (MIPS) have also revealed the presence of highly obscured, dusty star forming galaxies, previously undetected by optical or UV surveys \citep[e.g.][]{cedres2009,smith2010b}.   While the sensitivity of MIPS has enabled detailed studies of obscured star formation in individual local and distant galaxy clusters \citep[e.g.][]{geach2006,marcillac2007,dressler2009,haines2009,chung2010,finn2010}, there are still only a small number of low redshift clusters that have been systematically surveyed for dusty star-forming galaxies out to the virial radius.

There remain many uncertainties in the relationship between star formation in clusters and their global cluster properties.  In particular, several studies have tried to understand the correlation between cluster mass and the mass-normalized cluster star formation rate (SFR).  While results from \citet{bai2009} suggest that there is no strong correlation between cluster specific SFR and cluster mass, others such as \citet{finn2005}, \citet{poggianti2006}, and \citet{koyama2010} argue that cluster specific SFR decreases with cluster mass.

The large spatial coverage required to observe dusty star-forming galaxies in low redshift clusters out to the cluster infall regions has thus far hindered our ability to understand how star formation is affected by global cluster properties such as cluster mass.  In this paper we exploit data from the Wide-field Infrared Survey Explorer \cite[WISE;][]{wright2010} to overcome this observational challenge and present results on obscured star formation and how it relates to cluster mass and radius out to 3\rtwo\ in a sample of 69 clusters at $z<0.1$.   \rtwo\ and \mtwo\ are commonly used interchangeably with virial radius and total cluster mass, respectively.  \rtwo\ is the radius within which the average density is 200 times the critical density of the universe and \mtwo\ is the mass enclosed within that radius.

\section{Data} 
\subsection{WISE}

WISE is a medium-class explorer mission funded by NASA and has completed observations of the entire sky in four infrared bands: 3.4, 4.6, 12, and 22\micron\ (W1 to W4, respectively).   WISE scanned the sky with 8.8 second exposures, each with a 47 arcmin field of view, providing at least eight exposures per position on the ecliptic and increasing depth towards the ecliptic poles.  The individual frames were combined into coadded images with a pixel scale of 1.375 arcsec per pixel.  Cosmic-rays and other transient features were removed via outlier pixel rejection.

The photometry used for our analyses is point spread function (PSF) fitted magnitudes from the ``first-pass operations coadd source working database'' created by the WISE data reduction pipeline.  Galaxies in our cluster sample have a diffraction limited resolution of $12^{\prime\prime}$ (full width half maximum) in the 22$\mu$m band.  We have confirmed from W4 coadded images that all star-forming galaxies considered in our analyses appear unresolved in the 22\micron\ band, and have PSF photometry reduced $\chi^{2}$ values less than 1.5.   Therefore we use the PSF magnitudes from the first-pass photometric catalog to obtain estimates of total flux.

For the minimum coverage of 8 overlapping frames, the sensitivity for $S/N=5$ in the W4 band is 6 mJy, including uncertainty due to source confusion \citep{wright2010}.  To ensure an unbiased comparison of global SFRs and total IR luminosities of clusters at different redshifts, we impose a lower limit of SFR=4.6 \Msolaryr\ on our entire cluster sample, which is equivalent to a total IR luminosity of \LIR$>4.7\times10^{10} \Lsun$, and corresponds to the 6 mJy flux limit at $z=0.1$.   We hereafter refer to our sample of star-forming galaxies as demi-LIRGs, which have nearly half the total IR luminosity of a luminous infrared galaxy or LIRG.  However we note for future extragalactic studies using WISE data that most coadded observations will have at least 12 overlapping frames and hence better sensitivity than the conservative 6 mJy limit we adopt in this paper.  Additional information regarding WISE data processing is available from the Preliminary Data Release Explanatory Supplement. \footnote[1]{http://wise2.ipac.caltech.edu/docs/release/prelim/index.html}

\subsection{Cluster Sample}

We use the Cluster Infall Regions (CIRS; Rines \& Diaferio 2006) sample because it provides high-fidelity mass estimates, is at sufficiently low redshift to enable detection with WISE of strongly star-forming galaxies, and has extensive spectroscopy for membership determination.  The CIRS sample consists of 72 low-redshift X-ray galaxy clusters identified from the ROSAT All-Sky Survey  that are within the spectroscopic footprint of SDSS Data Release 4.  The redshift range of the CIRS clusters is $0.003<z<0.1$, with a median of $z=0.06$.  Cluster masses are available from \citet{rines2006}, who utilize the caustics infall pattern to determine total dynamical cluster mass and \rtwo\ \citep{diaferio2005}.  The clusters in this paper consist of the entire CIRS sample, excluding three clusters at $z\leq0.006$,  which leaves 69 remaining clusters with a minimum redshift of $z=0.0204$. 

\subsection{SDSS DR7}

Optical photometric and spectroscopic data are obtained from the Sloan Digital Sky Survey Data Release 7 (SDSS DR7) \citep{abazajian2009}, which are 90\% spectroscopically complete for galaxies with $r<17.77$ and half-light surface brightness $\mu_{50}<24.5$ mag arcsec$^{-2}$.  However, the spectroscopic completeness is lower in high-density regions such as in the core of galaxy clusters, due to fiber collisions.  Adjacent fibers cannot be placed closer than 55 arcsec from each other, which corresponds to a separation of 63 kpc at $z=0.06$.  To verify that the spectroscopic incompleteness arising from constraints on fiber placement has a negligible effect on radial trends of star formation in clusters, we look at the fraction of W4-bright sources as a function of distance from the cluster center.

Among the  $r<17.77$ sample, the spectroscopic completeness of W4-bright sources (W4 $S/N>5$) is $\sim$70\% within the central 5 arcmin radial bin, and increases to $\sim$80\% in the 5-10 arcmin bin.   In other words, we are not missing a significant fraction of star-forming galaxies in the cluster core due to fiber collisions.  The spectroscopic completeness for all galaxies (regardless of IR emission) with $r<17.77$  shows a similar behavior with distance from the cluster center, and on average reaches at least $\sim$80\% completeness beyond the central 5 arcmin radius.  Since the spectroscopic incompleteness of both W4-bright and optically bright galaxies reaches $\sim$80\% beyond the cluster core, our results on radial trends of star formation are negligibly affected by spectroscopic incompleteness.  In addition, we are measuring trends on scales of hundreds of kpc, and therefore not significantly affected by small differences in spectroscopic completeness near the cluster core.

We use photometric data from SDSS DR7 to estimate stellar masses for cluster members, using the tight correlation between stellar mass-to-light (M/L) ratio and optical colors, as determined by \citet{bell2003}.  With a sample of more than 10,000 optically bright galaxies ($13<r<17.5$), \citet{bell2003} construct a grid of stellar population models with a range of metallicities and star formation histories, then compare the best-fit galaxy templates with evolved zero redshift templates to determine present-day M/L ratios.    We use the relation between $g-r$ color and $r$-band stellar M/L ratio to derive stellar masses for our sample.   The estimated uncertainty of the color-based stellar M/L ratios, including random and systematic errors, is $\sim$45\% \citep{bell2003}.

\section{Analysis}
\subsection{Cluster Membership}

\begin{figure}[t]
\epsscale{1.15}
\plotone{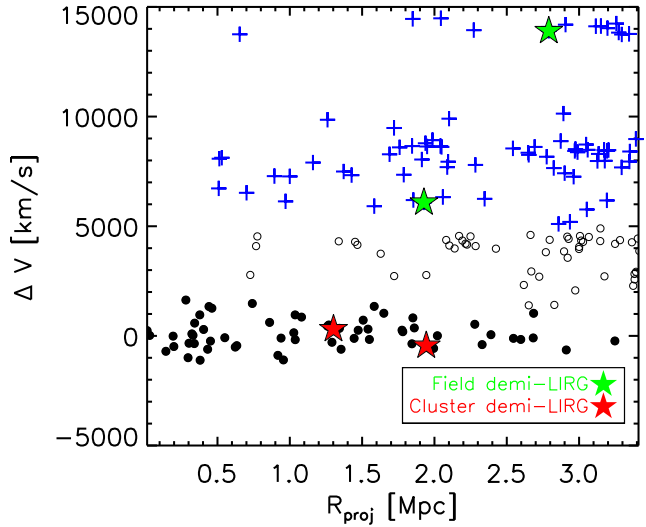}
\caption{Velocity offset versus projected distance from the cluster center of Abell 1377, for all SDSS galaxies brighter than  $r<17.77$ with spectroscopic redshifts, within a projected 5 Mpc radius of the cluster center.  Filled circles indicate the cluster members within 3\rtwo.  Cross symbols (blue) highlight field galaxies that are at least 5000 \kms\ away from the cluster systemic velocity.  Red and green star symbols represent  demi-LIRGs in the cluster and field populations, respectively.  Open circles are galaxies that are included in the SDSS spectroscopic sample but are not chosen as cluster members nor field galaxies.}
\label{fig:caustic}
\end{figure}

Spectroscopic redshifts from SDSS DR7 are used to determine membership for each of our clusters.  Figure~\ref{fig:caustic} illustrates an example of using the caustics infall method to determine cluster membership for Abell 1377, the most massive cluster in our sample.  It shows the difference in radial velocity ($cz$) of galaxies in the cluster field with respect to the cluster systemic velocity, as a function of projected distance from the cluster center.   Galaxies that are dynamically bound to the cluster form a well-defined region that decreases in velocity offset as a function of projected radius.   We define the galaxies that are within the edge of this envelope and within a projected radius of $\leq$3\rtwo, as cluster members.   While the average turnaround radius for the CIRS sample is $\sim$5\rtwo, we restrict our cluster galaxy sample to within 3\rtwo, since  cluster infall patterns are generally better defined closer in to the cluster center. 

Only galaxies with SDSS spectroscopic redshifts are considered in the following analyses.  We also limit all cluster and field galaxies to be brighter than $M_{r}=-20.3$, which corresponds to the 90\% spectroscopic completeness limit of $r=17.77$ at $z=0.1$.

\subsection{Star Formation Rates and Infrared Luminosities}

Star formation rates and total infrared luminosities (\LIR) are determined from the WISE 22\micron\ photometry, using the relations presented in \citet{rieke2009} that are calibrated for MIPS 24\micron\ data.  \citet{rieke2009} constructed model average spectral energy distribution (SED) templates from a sample of local LIRGs and ULIRGs and derive correlations between 24\micron\ flux, SFR, and \LIR.  The flux-SFR relation given in their equation~14 can be used for both MIPS 24\micron\ and WISE 22\micron\ data \citep{rieke2009}.

The similarity of the two bandpasses are confirmed by \citet{goto2011} who quantify the correlation between \LIR\ and the inferred MIPS 24\micron\ and WISE 22\micron\ luminosities.  The MIPS and WISE luminosities are inferred by applying a color-correction to the 18\micron\ AKARI flux, and \LIR\ is derived from fitting the IR SED templates of \citet{chary2001} to all six AKARI bandpasses (9, 18, 65, 90, 140, and 160 \micron) for $\sim$600 galaxies at $z<0.1$.  The resulting correlations between the $\mathrm{L_{IR}-L_{MIPS24}}$ and $\mathrm{L_{IR}-L_{WISE22}}$ are nearly identical, with only a $\sim$4\% offset.  Therefore we proceed by using the flux-SFR and $\mathrm{L_{IR}-L_{MIPS24}}$ relations of \citet{rieke2009}, assuming the calibration determined from MIPS 24\micron\ data.

The total error attributed to the SFR is $\sim$0.2 dex, which is dominated by scatter in the $L_{MIPS24}$-\LIR\ relation, and does not include uncertainties inherent in the assumed stellar initial mass function (IMF).  The IMF adopted by \citet{rieke2009} is similar to the \citet{kroupa2002} and \citet{chabrier2003} IMFs, which have relatively fewer low mass stars compared to a Salpeter IMF, and is more applicable for extragalactic star-forming regions \citep{rieke1993,alonso2001}.

\subsection{Exclusion of AGN}

\begin{figure}[t]
\epsscale{1.15}
\plotone{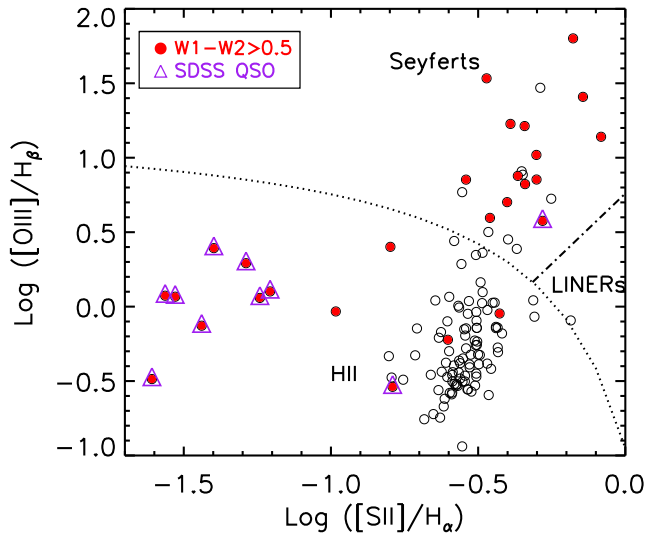}
\caption{The BPT diagram for 136 cluster galaxies with $L_{IR}>4.7\times10^{10}$ \Lsun.  Dotted curves indicate boundaries from \citet{kewley2006} that separate galaxies whose emission lines originate from Seyferts, LINERs, and HII regions.  Solid circles (red) highlight galaxies that are identified as AGN based on having a red WISE color (W1-W2$>$0.5), and triangles (purple) represent sources flagged as quasars in the SDSS catalog.}
\label{fig:BPT}
\end{figure}

A robust estimate of the cluster star formation rates requires that we first identify and exclude all 22$\mu$m sources that are dominated by AGN rather than star formation. As demonstrated in \citet{chung2010}, even a few IR-bright AGN can significantly bias the inferred global star formation rate of a cluster. There are several methods by which we can identify AGN in our data set. The first method relies upon WISE colors. Similar to the AGN wedge in Spitzer/IRAC data \citep{stern2005}, AGN-dominated sources will have red colors in W1-W2 ([3.4]-[4.6]).  Specifically, we use the criteria W1-W2$>$0.5 (Vega) to identify candidate AGN in the WISE data set.  This color selection is similar to, though slightly bluer than, the AGN selection determined in Jarrett et al. 2011 (submitted) and Stern et al. 2011 (in prep).   While this single color cut is in general less robust than the full AGN wedge, at the low redshifts that are the focus of this work, star-forming galaxies should have colors uniformly blueward of this threshold.  We illustrate in Figure~\ref{fig:BPT} that our W1-W2$>$0.5 criterion works well in selecting out AGN from star-forming galaxies.   Profile-fit photometry is used to determine the W1-W2 colors and AGN exclusion.  While some of the nearest galaxies will be resolved in these bands, we are interested only in the W1-W2 color, rather than in single-band photometry.  Comparison of AGN selection based on W1-W2 color from aperture photometry and profile-fit photometry show that the two methods are similarly successful in isolating AGN.  However, the color cut based on profile-fit magnitudes has a slightly better overlap with the optically detected AGN, discussed below.

A second approach to AGN identification is use of optical spectroscopy to identify sources that lie in the AGN region of the Baldwin-Phillips-Terlevich (BPT) diagram  \cite[BPT;][]{baldwin1981}. Here we use the BPT diagram as a cross-check on the WISE color selection for the subset of sources.  Figure~\ref{fig:BPT} shows the emission line ratios of [OIII]/H$_{\beta}$ and [SII]/H$_{\alpha}$, for a sample of 136 cluster members with $L_{IR}>4.7\times10^{10}$ \Lsun, of which 27 have W1-W2$>$0.5.  Boundaries from \citet{kewley2006} are shown as dotted lines and separate regions where narrow emission lines arise from the presence of Seyferts, LINERs (low ionization narrow emission line regions), and HII regions (star-forming galaxies).  Galaxies that are selected as AGN candidates based purely on having a red WISE color (W1-W2$>$0.5) are indicated as filled (red) circles.

Figure~\ref{fig:BPT}  shows that out of 22 optically identified Seyferts, fourteen (65\%) are also identified as AGN based on the WISE W1-W2$>$0.5 selection.   Sources flagged as quasars (or QSOs) in the SDSS catalog are highlighted with (purple) triangle symbols, with all ten quasars being independently identified as AGN using the WISE color selection.   Since the BPT diagnostic is meant to classify galaxies based only on {\it narrow} emission line ratios, it is not surprising that nearly all of the SDSS quasars are not properly diagnosed in the BPT diagram.  The H$\alpha$ and H$\beta$ emission lines in eight out of the ten SDSS quasars have broad wings relative to their neighboring [OIII] or [SII] emission lines.  Overall, we conclude from Figure~\ref{fig:BPT} that the WISE color selection is an effective method of excluding AGN from our sample of bright W4 sources, identifying 85\% of the optically detected Seyferts and quasars.

\section{Results \& Discussion} 

The two central questions that we aim to address are how the mean specific star formation rate of cluster galaxies (mSSFR) depends upon location within a cluster, and how the total integrated star formation rate of a cluster depends upon cluster mass.  Specifically, the mSSFR is defined as the total star formation rate in cluster galaxies {\it at} a given projected radius as inferred from 22$\mu$m photometry, divided by the total stellar mass of all cluster galaxies {\it at} that radius.

For probing the total integrated star formation rate we also define the integrated cluster quantity
\begin{equation}
cSSFR= \frac{SFR (r<3R_{200})}{M_{200}},
\end{equation}
which is a useful mass-normalized measure of the total star formation rate within the infall region, and referred to as the cluster specific SFR (cSSFR).

Among the 69 clusters, a total of 136 demi-LIRGs are detected within 3\rtwo, of which 27 are determined to be AGN based on their W1-W2 color.   In the following sections, all SFR quantities are determined from the remaining 109 star-forming demi-LIRGs.  Eight Seyferts identified from the BPT diagram are included in the cluster star-forming galaxy sample because their W1-W2 colors are not indicative of AGN activity, and we prefer to maintain a uniform WISE selection of AGN among all field and cluster galaxies.  We note that all results presented in this paper are negligibly affected by the exclusion/inclusion of these eight Seyferts.  The coordinates, redshift, W4 magnitude, SFR, \LIR, projected distance from cluster center, and AGN flags are listed for the 136 demi-LIRGs in Table~1.

\subsection{Radial Dependence of Star Formation}

\begin{figure}[t]
\epsscale{1.15}
\plotone{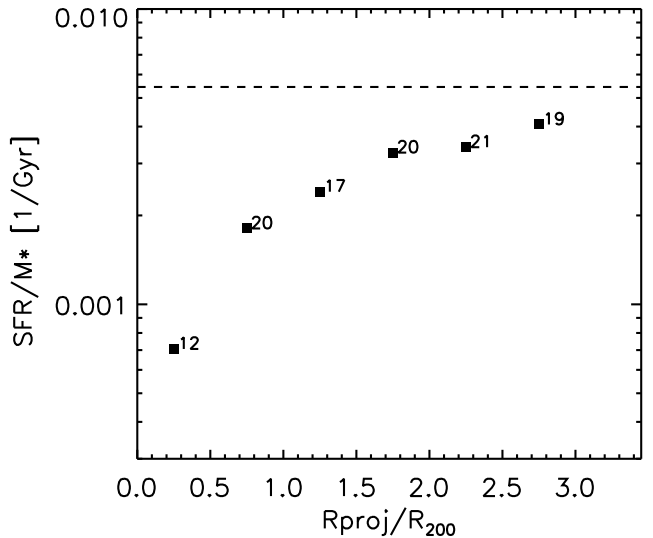}
\caption{Mean specific SFR (mSSFR) of spectroscopically confirmed cluster members
  as a function of projected radius, with numbers indicating how many
  star forming galaxies with $L_{IR}>4.7\times10^{10}\Lsun$\ are in each
bin.  Dashed line indicates the mSSFR of the field sample.}
\label{fig:SSFR_radius}
\end{figure}

There is a long history in the literature demonstrating the existence of a strong radial dependence for star formation (or color) in galaxy clusters \citep[e.g.][]{lewis2002,gomez2003,balogh2004}. The main strengths of the current analysis are the uniform 22$\mu$m WISE data, spectroscopic completeness, and existence of \rtwo\ measurements for the full sample, which together provide us with an infrared-based view of star formation for a homogeneous sample extending well beyond the virial radius.  A common approach in the literature has been to look at the radial dependence of the fraction of star-forming galaxies.  Here we investigate both the dependence of the star-forming fraction, and also the mSSFR out to 3\rtwo.

For comparison, the identical quantities for a field  population are calculated, with the field sample chosen from a catalog of galaxies located within a projected radius of 5 Mpc from the cluster center, at the cluster redshift.  Among these galaxies, we choose those with a radial velocity greater than 5000 \kms\ away from the cluster systemic velocity, redshift $z<0.1$, and absolute r magnitude $M_{r}<-20.3$.  These are the same redshift and magnitude limits of the cluster galaxy sample.   The highest velocity dispersion of our cluster sample is $\sim$960 \kms, which means that the chosen field galaxies are more than $5\sigma$ away from the cluster redshift.  Figure~\ref{fig:caustic} shows delta-velocity ($cz$) versus projected distance from the cluster center for galaxies within a 5 Mpc projected radius of Abell 119.  The field galaxies indicated with (blue) cross symbols are clearly not associated with the cluster galaxies, which appear distinctly confined to a trumpet-shaped region.  By gathering field galaxies within a projected 5 Mpc radius from 69 different regions of the sky, we have compiled a large enough sample to obtain a representative field value of mSSFR and the fraction of star-forming galaxies with $L_{IR}>4.7\times10^{10}$ \Lsun.  There are a total of 11180 field galaxies, of which 566 are demi-LIRGs.

Figure~\ref{fig:SSFR_radius} shows the mSSFR as a function of r/\rtwo\ for our ensemble of 69 clusters, including star formation only for galaxies  with $L_{IR}>4.7\times10^{10}$ L$_\odot$ (SFR$>4.6$ M$_\odot$ yr$^{-1}$).   We emphasize that by construction our SFR limit therefore means that the observed mSSFR is a lower bound -- but a consistent lower bound across the sample.

\begin{figure}[t]
\epsscale{1.15}
\plotone{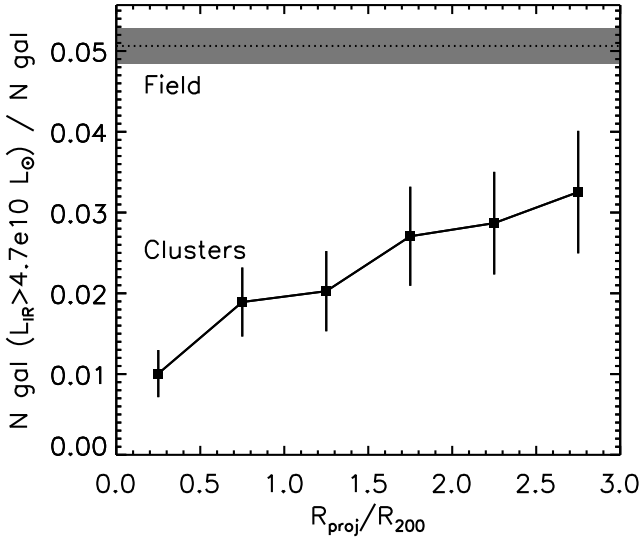}
\caption{Ratio of star-forming galaxies to all spectroscopic
  cluster members as a function of projected cluster radius, with the
  typical field value shown as a dashed line.}
\label{fig:lirg_fraction}
\end{figure}

As expected, the mean specific SFR increases with projected radius, with the  most central bin containing the fewest star-forming demi-LIRGs -- twelve out of a total of 109 demi-LIRGs found in 69 clusters.  The mSSFR displays a steep increase from the central bin to \rtwo, then continues to increase monotonically and nearly flattens out below the field value at larger radii.  This is consistent with a low redshift study of H$\alpha$  star-forming  galaxies by \citet{lewis2002}, who found that the median SFR (normalized by L$^{*}$) of cluster galaxies reaches the field value beyond 3\rtwo.

 Such a radial trend can be driven by two factors -- an increase in the SSFR of the sub-population  of star-forming galaxies, or an increase in the fraction of star-forming galaxies with radius.  We find that there is no statistically significant change in the SSFR of individual star-forming galaxies with radius, implying that the trend is driven primarily by a radial gradient in the star forming fraction.

This can be seen in Figure~\ref{fig:lirg_fraction}, where we directly plot the fraction of star-forming galaxies as a function of radius.  More than one third of the star-forming galaxies reside at 2\rtwo$<$r$<$3\rtwo.   What is striking  is that even at these large radii the mSSFR and star-forming fraction remain below the field values.  In Figure~\ref{fig:lirg_fraction}, the demi-LIRG fraction is higher in the field relative to the cluster population at 3\rtwo\ by a factor of $\sim$1.5.  One possible interpretation of this result is that these are infalling galaxies that may already be ``pre-processed'' in intermediate-density environments, such as galaxy groups or filaments (e.g. Zabludoff \& Mulchaey 1998), which would explain why they are suppressed relative to a true field population.

There is both observational and theoretical support in favor of the suppressed star-forming galaxy fraction out to 3\rtwo.  \citet{balogh1998} found that the mean cluster SFR, which they derived from [OII] emission of galaxies in 15 clusters, is lower than the mean field SFR by more than a factor of two at 2\rtwo.  Results from \citet{lewis2002} and \citet{linden2010} similarly conclude that cluster star formation is suppressed at several times the virial radius relative to the field.  

In addition to galaxies being pre-processed at large radii, another explanation for the suppressed demi-LIRG fraction in Figure~\ref{fig:lirg_fraction} could be the contribution of a ``backsplash'' population beyond \rtwo.  Simulations from \citet{balogh2000} and \citet{gill2005} show that up to $\sim$50\% of galaxies currently located between \rtwo\ and 2\rtwo\ may have previously traveled inward of the virial radius of the cluster.  Therefore the suppressed demi-LIRG fraction at large radii may, to some degree, reflect quenched star formation that occurred within the cluster core.   However our galaxy sample, which has an absolute magnitude limit of $M_{r}<-20.3$, is sensitive to massive galaxies brighter than M$^{*}+1$ \citet{blanton2003}, whereas backsplash galaxies are on average significantly less massive than first infall galaxies in the cluster outskirts \citep{gill2005}. As such, our sample is biased against the lower mass backsplash galaxies relative to first infall galaxies.  Nonetheless, a contribution of backsplash galaxies may indeed be partially responsible for the suppressed star formation activity at large distances from the cluster core. 


\subsection{Cluster Specific SFR versus Cluster Mass}

Total mass is, both for astrophysics and cosmology, the most fundamental physical galaxy cluster parameter. There are multiple physical processes that depend on the depth of the cluster potential well (e.g. ram pressure, harassment, tidal interactions) which can significantly alter the morphologies, gas content, and star formation rates in cluster galaxies \citep[e.g.][]{gavazzi2001,bekki2002,owen2005,haynes2007}.  Furthermore, it has been observed that while the total baryon fraction (stellar and gas content) remains roughly constant with cluster mass, the stellar mass decreases and gas content increases in galaxy clusters as a function of cluster mass \citep{gonzalez2007,giodini2009,andreon2010}.  This implies that the integrated star formation efficiency of the history of a cluster is directly tied to cluster mass.

In this section we examine the relation between cluster mass and total current cluster star formation to better understand how strongly the present-day star formation rate depends upon cluster mass.  Our approach is to compute the cluster SFR normalized by cluster mass (cSSFR; equation~1) as a function of cluster mass.  For each cluster, the cluster SFR is the sum of SFRs for all member galaxies that have SFR$>$4.6 \Msolaryr\ and lie within a projected radius of 3\rtwo.

\begin{figure*}[t]
\epsscale{1}
\plottwo{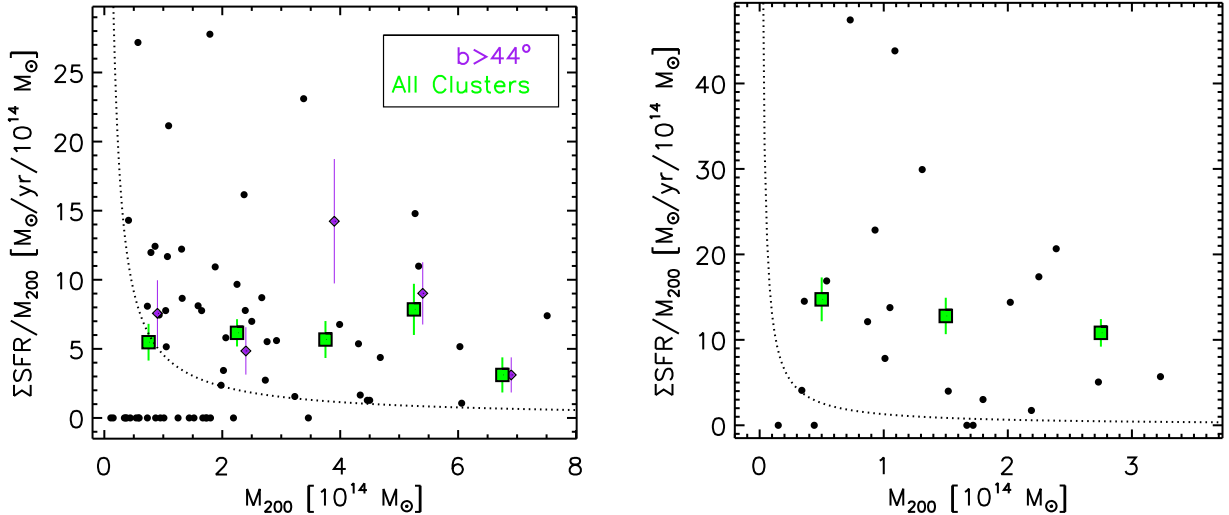}{fig5b.epsi}
\caption{Left: Cluster specific SFR versus
  cluster mass for 62 clusters, with the average value per mass bin overplotted with large green square symbols.  The average cluster specific SFR calculated using only clusters at high ecliptic latitude (and therefore greater coverage with WISE) are shown as purple diamonds, offset from the green square symbols for clarity.  The subsample of  clusters at high ecliptic latitude shows a similar trend relative to the full sample of clusters, indicating that the gaps present in the WISE coadd data at low ecliptic latitudes do not have a significant impact on our result. The dotted line shows the cluster specific SFR when SFR=4.6 \Msolaryr\, corresponding to the 6 mJy W4-band detection threshold at $z=0.1$.  Right: Cluster specific SFR versus cluster mass for a low redshift subsample of 22 clusters at $z<0.06$.  The dotted line shows a detection limit of SFR=1.4 \Msolaryr, corresponding to the 6 mJy W4-band detection threshold at $z=0.06$.  There is no apparent trend between cluster specific SFR and \mtwo\ even with a lowered SFR threshold.}
\label{fig:SSFR_M200}
\end{figure*}

The left panel of Figure~\ref{fig:SSFR_M200} shows the cSSFR for 62 clusters, which consists of the sample of 69 clusters, excluding seven clusters that have incomplete spatial coverage in the SDSS DR7 spectroscopic survey within 3\rtwo.  The dotted curve shows the limiting detectable cSSFR as a function of \mtwo\ corresponding to a single demi-LIRG with a SFR of 4.6 \Msolaryr\ in a cluster.  Clusters that have no members above the SFR$>$4.6 \Msolaryr\ limit are assigned a cSSFR of zero.  The binned data (large squares) indicate the average cSSFR in each \mtwo\ bin, including clusters with a cSSFR of zero.  Thus the cSSFR of the lowest \mtwo\ bin is strongly biased by incompleteness, since nearly all of the clusters at low mass fall below the SFR$>$4.6 \Msolaryr\ limit and have a cluster SFR set to zero.  To ensure that our results are not significantly biased by this incompleteness at low cluster mass, we also show the cluster specific SFR versus \mtwo\ for a sub-sample of clusters at $z<0.06$ in the right panel of Figure~\ref{fig:SSFR_M200}.  For these 22 low-z clusters, we are sensitive to star-forming galaxies down to SFR$>$1.4 \Msolaryr, which corresponds to the W4 band 5$\sigma$ detection limit of 6 mJy at $z=0.06$, and is illustrated with a dotted curve.

Using the sample of 62 clusters, the left panel of Figure~\ref{fig:SSFR_M200} shows no significant correlation between cluster specific SFR and cluster mass, over more than one order of magnitude in $M_{200}$.  The first pass coadd processing used here has the same calibration as in the WISE Preliminary Data release, but was run every other day using only one day of data, creating gaps in coverage at low ecliptic latitude.  These gaps should not significantly impact this result, since location of the gaps are random with respect to the cluster centers.  We confirm this expectation by examining the cluster specific SFRs of a sub-sample of 24 clusters with ecliptic latitude $b>44$ deg.  These clusters are at sufficiently high ecliptic latitude such that they have fairly complete spatial coverage with the first-pass WISE coadd data.  The average cluster specific SFRs of the high ecliptic latitude sub-sample are overplotted in Figure~\ref{fig:SSFR_M200} in small purple diamonds.  These high ecliptic latitude clusters show no significant correlation between cluster specific SFR and cluster mass, similar to the lack of correlation seen with the full cluster sample.

The lack of correlation between cSSFR and \mtwo\ is also confirmed with the $z<0.06$ sub-sample, for which we have applied a significantly lower detection limit of SFR$>$1.4 \Msolaryr\ (Figure~\ref{fig:SSFR_M200}, right panel).   With the lower SFR limit, the  clusters in Figure~\ref{fig:SSFR_M200} have an average of 5.2 star-forming galaxies within R$<$3\rtwo, with a large scatter.  Five of the low redshift clusters have between 10 to 16 star-forming galaxies.  In contrast, for the  sample of 62 clusters with the demi-LIRG cut (SFR$>$4.7 \Msolaryr), there is an average of 1.6 star-forming galaxies per cluster within R$<$3\rtwo.  While the average number of star-forming galaxies per cluster is small in this case, the cSSFRs calculated using only the demi-LIRGs are well correlated with the cSSFRs using the SFR$>$1.4 \Msolaryr\ cut, with a linear Pearson correlation coefficient of 0.9.  Therefore, even with an average of only 1.6 star-forming galaxies per cluster, the cSSFRs of the full cluster sample are qualitatively representative of cSSFRs that sample the fainter end of the infrared luminosity function.

 The binned data in both panels of Figure~\ref{fig:SSFR_M200} show cluster specific SFRs that are consistent within 1$\sigma$ of each other, across the full range of \mtwo.  The error bars associated with the average cSSFR are the Poisson uncertainty on the number of star-forming galaxies detected above the SFR limit in each \mtwo\ bin.  The lack of a significant trend between cSSFR and \mtwo\ indicates that transformation mechanisms which scale strongly with cluster mass may not play a dominant role in the evolution of star formation in clusters.

While Figure~\ref{fig:SSFR_M200} illustrates the integrated cluster SFR per \mtwo\ by taking the sum of all star-forming galaxies  within R$<$3\rtwo, our results remain robust when considering only those galaxies within R$<$\rtwo.  Since galaxies within \rtwo\ are more susceptible to global cluster processes such as ram pressure, the lack of correlation between cSSFR (R$<$\rtwo) and cluster mass re-affirms the main conclusion drawn from Figure~\ref{fig:SSFR_M200} -- mechanisms that are strongly dependent on cluster mass do not play a significant role in the evolution of star formation in clusters.

Two cluster processes that strongly scale with cluster mass are galaxy harassment and ram pressure.  Ram pressure is a mechanism that acts to compress and/or strip cold gas reservoirs in galaxies.  The impact of ram pressure is proportional to $\rho v^{2}$, where $\rho$ is the density of the intracluster medium (ICM) and $v$ is the velocity of a galaxy with respect to the ICM \citep{gunn1972}.  Galaxy harassment refers to the cumulative effect of high velocity close encounters between galaxies and has the potential to disturb morphologies and quench star formation \citep{moore1996}.  The effects of galaxy harassment, like ram pressure, are expected to scale up with cluster mass, since both cluster velocity dispersion and ICM gas density increase with cluster mass.

While there has been evidence in support of both ram pressure and harassment having an impact on star formation in individual cluster galaxies \citep[e.g.][]{roediger2005,vollmer2008,haynes2007}, the lack of a correlation between cluster specific SFR and cluster mass in Figure~\ref{fig:SSFR_M200} indicates that neither of these mechanisms significantly trigger/quench star formation in local galaxy clusters, within the mass range of $10^{14}$ to $7\times10^{14}$ \Msun.   

Our results are consistent with those of \citet{goto2005} and \citet{popesso2007},  who use a sample of $\sim$100 clusters at $z<0.1$ and find no significant correlation between H$\alpha$-derived cluster specific SFR and cluster mass.  Both \citet{goto2005} and \citet{popesso2007} use SDSS DR2 data, with the same limiting r-band magnitude as used in this paper.  The limiting H$\alpha$ SFR is $\sim$3 \Msolaryr.  Similarly, \citet{balogh2010} demonstrate that the passive galaxy fraction is roughly constant as a function of system mass, from $M\sim10^{13}$ \Msun\ to $10^{15}$ \Msun.  Assuming that the non-passive galaxy fraction is dominated by star-forming galaxies rather than AGN, the results of \citet{balogh2010} support our finding that cluster specific SFR is not strongly dependent on cluster mass.    

In contrast, several other studies have found evidence for an anti-correlation between cluster specific SFR and cluster mass.   \citet{finn2005} and \citet{koyama2010} obtain cluster SFRs by integrating H$\alpha$ derived SFRs for members out to 0.5\rtwo, and cluster masses from observed velocity dispersions for a sample of eight and nine clusters, respectively. Their clusters span an approximate redshift range of 0.2 to 0.8, and the H$\alpha$ SFRs are sensitive to $\sim$1 \Msolaryr.   The anti-correlation observed between cluster specific SFR and cluster mass is significantly weakened when only clusters of the same epoch are considered, indicating that evolutionary effects may be the dominating factor.

In addition to interactions between  galaxies and the cluster environment, studies have demonstrated that tidal interactions between galaxies or galaxy-galaxy mergers can trigger a burst of star formation \citep[e.g.][]{owen2005,martig2008}.  Galaxy-galaxy interactions are generally optimized in intermediate density regions with low velocity dispersions, such as in galaxy groups, whereas the more massive clusters are less likely to host galaxy mergers.

Figure~\ref{fig:SSFR_M200} includes a substantial number of low mass systems (57 clusters with \mtwo$<3\times10^{14}$ \Msun), yet shows no sign of significantly elevated cluster specific SFR in low mass clusters relative to intermediate mass clusters.   Any signature of enhanced star formation in low mass clusters due to galaxy-galaxy interactions should be particularly evident in our data, because we integrate the SFR of cluster members out to three times the virial radius, where galaxy-galaxy mergers and tidal interactions are more like to occur.   A comparison of the distance to nearest neighbor among the demi-LIRG population and the general cluster population as a function of projected radius shows that the two populations have mean nearest neighbor distances that are consistent within 1$\sigma$.  This indicates that galaxy-galaxy mergers or close tidal interactions do not play a dominant role in the star formation properties of these clusters.

Our results imply that even in the low mass $\sim5\times10^{13}$ \Msun\ groups, star formation has already been quenched to similar levels observed in clusters that are more massive by over an order of magnitude.  The lack of correlation between cSSFR and \mtwo\ (Figure~\ref{fig:SSFR_M200}), combined with the suppressed demi-LIRG fraction at 3\rtwo\ relative to the field (Figure~\ref{fig:lirg_fraction}, would suggest a scenario in which star formation is quenched in a significant population of galaxies within small groups or filaments.  Then once the remaining star-forming galaxies are accreted into the cluster environment, a mechanism that is not strongly dependent on cluster mass must operate.  We suggest strangulation as a plausible candidate for quenching star formation in cluster star-forming galaxies.  

As a galaxy enters the cluster ICM, its hot halo gas is removed, thereby cutting off its resource for future cold gas supplies, since the halo gas would otherwise eventually cool and settle onto the disk \citep{balogh2000,bekki2002}.  The effectiveness of strangulation is less dependent on cluster mass than ram pressure or harassment.  Simulations from \citet{kawata2008} have shown that strangulation can occur in low mass groups, where the ICM-galaxy halo interaction is weak relative to clusters.   Strangulation has been suggested by several cluster studies as an important cluster mechanism \cite[e.g.][]{moran2006,moran2007,chung2010}, including by \citet{treu2003} who discovered ``mild gradients'' in the morphological fractions of galaxies in Cl0024+16 out to large cluster radii, which were best explained by a slow-working gentle mechanism such as strangulation.   

While strangulation may quench star formation in massive clusters, our results suggest that the dominant process for regulating star formation  in dense environments does not depend upon cluster mass, and hence must be efficient in small groups or filaments.

\section{Conclusions} 

We use data from WISE and SDSS DR7 to study the dependence of star formation on cluster mass and location within a cluster in 69 local ($z<0.1$) galaxy clusters.  Cluster membership is determined from SDSS DR7 spectroscopic redshifts for galaxies brighter than $M_{r}=-20.3$, and star formation rates are determined with 22\micron\ photometry from WISE using the relations outlined in \citet{rieke2009}.

Out of 69 clusters, we find a total of 109 star-forming demi-LIRGs with SFR$>4.6$ \Msolaryr within 3\rtwo.  Both the fraction of demi-LIRGs and the mean specific SFR of cluster galaxies increases with projected distance from the cluster center.  However, the fraction of demi-LIRGs remains below the field value even at 3 times the cluster virial radius.  One plausible explanation for the suppressed demi-LIRG fraction at 3\rtwo\ is that even galaxies that reside significantly beyond the virial radius  have already been quenched in their previous environments, such as small groups or filaments.

We also investigate the impact of cluster mass on star formation by presenting the total cluster SFR normalized by \mtwo\ as a function of \mtwo.  We find no evidence of a correlation between the cluster specific SFR and cluster mass in this first uniform dataset to detect obscured star formation out to several times the virial radius in a large sample of low redshift clusters.  Our result indicates that cluster mechanisms which scale with cluster mass, such as ram pressure or harassment, are not likely to play a dominant role in the evolution of star formation in local clusters.

\acknowledgements This publication makes use of data products from the Wide-field Infrared Survey Explorer, which is a joint project of the University of California, Los Angeles, and the Jet Propulsion Laboratory/California Institute of Technology, funded by the National Aeronautics and Space Administration.  The authors would like to thank Emilio Donoso for his help and advice on navigating the SDSS database.  We also thank the anonymous referee for a careful reading and comments which improved the paper.


\clearpage
\newpage 

\scriptsize
\begin{center}
\begin{longtable}{p{1.7in}p{0.5in}p{0.5in}p{0.3in}p{0.25in}p{0.25in}p{0.3in}p{0.5in}p{0.5in}p{0.5in}}

\caption{List of demi-LIRGs from 69 clusters sorted in descending order of \LIR.  Galaxies marked as AGN have WISE color W1-W2$>$0.5 and are assumed to have a mid-IR flux dominated by an AGN rather than star formation.  The six BPT Seyferts with W1-W2 color consistent with star formation rather than AGN, are marked with an asterisk.} \\ 

\hline \hline \\
\multicolumn{1}{c}{\textbf{WISE Name}}  & \multicolumn{1}{c}{\textbf{RA [deg]}}  & \multicolumn{1}{c}{\textbf{Dec [deg]}} & \multicolumn{1}{c}{\textbf{z}} & \multicolumn{1}{c}{\textbf{W4}} & \multicolumn{1}{c}{\textbf{W4 sig}} & \multicolumn{1}{c}{\textbf{SFR}} & \multicolumn{1}{c}{\textbf{L$_{IR}$ [\Lsun]}}  & \multicolumn{1}{c}{\textbf{R$\mathrm{_{proj}}$ [\rtwo]}}  & \multicolumn{1}{c}{\textbf{AGN Flag}} \\

\hline
\endfirsthead

\multicolumn{10}{c}{{\tablename} \thetable{} -- Continued} \\
\hline \hline \\
\multicolumn{1}{c}{\textbf{WISE Name}}  & \multicolumn{1}{c}{\textbf{RA [deg]}}  & \multicolumn{1}{c}{\textbf{Dec [deg]}} & \multicolumn{1}{c}{\textbf{z}} & \multicolumn{1}{c}{\textbf{W4}} & \multicolumn{1}{c}{\textbf{W4 sig}}  & \multicolumn{1}{c}{\textbf{SFR}} & \multicolumn{1}{c}{\textbf{L$_{IR}$ [\Lsun]}}  & \multicolumn{1}{c}{\textbf{R$\mathrm{_{proj}}$ [\rtwo]}}  & \multicolumn{1}{c}{\textbf{AGN Flag}} \\
\hline
\endhead

\multicolumn{10}{c}{{Continued on Next Page\ldots}} \\
\endfoot

\hline \hline
\endlastfoot

WISEPC J235654.30-101605.2 &  359.22626 &  -10.26812 & 0.074 & 4.12 & 0.025 &  90.3 & 7.2e+11 & 1.1 & AGN \\ 
WISEPC J155850.42+272324.5 &  239.71010 &   27.39013 & 0.093 & 4.76 & 0.024 &  86.7 & 7.0e+11 & 1.7 & AGN \\ 
WISEPC J121635.79+040709.2 &  184.14912 &    4.11922 & 0.076 & 4.62 & 0.027 &  58.5 & 4.9e+11 & 2.1 & AGN \\ 
WISEPC J130534.25-021119.1 &  196.39270 &   -2.18864 & 0.088 & 5.41 & 0.039 &  38.8 & 3.4e+11 & 0.2 & AGN \\ 
WISEPC J121742.00+034631.2 &  184.42499 &    3.77535 & 0.080 & 5.25 & 0.032 &  35.6 & 3.1e+11 & 1.8 & AGN \\ 
WISEPC J160003.13+263707.8 &  240.01305 &   26.61884 & 0.093 & 5.67 & 0.033 &  35.1 & 3.1e+11 & 2.0 & AGN \\ 
WISEPC J111519.95+542316.6 &  168.83313 &   54.38794 & 0.070 & 4.98 & 0.026 &  34.2 & 3.0e+11 & 1.5 & AGN \\ 
WISEPC J171031.49+643914.6 &  257.63123 &   64.65405 & 0.079 & 5.41 & 0.026 &  29.7 & 2.6e+11 & 0.3 & SF \\ 
WISEPC J134236.22+592324.7 &  205.65091 &   59.39019 & 0.071 & 5.15 & 0.030 &  29.7 & 2.6e+11 & 0.5 & AGN \\ 
WISEPC J170226.20+341117.5 &  255.60916 &   34.18821 & 0.105 & 6.16 & 0.035 &  28.9 & 2.6e+11 & 0.1 & AGN \\ 
WISEPC J111448.38+401749.3 &  168.70160 &   40.29702 & 0.076 & 5.47 & 0.032 &  25.8 & 2.3e+11 & 1.2 & SF* \\ 
WISEPC J162021.16+301020.7 &  245.08818 &   30.17243 & 0.096 & 6.14 & 0.037 &  23.7 & 2.2e+11 & 2.6 & AGN \\ 
WISEPC J160105.12+272539.2 &  240.27135 &   27.42756 & 0.087 & 5.91 & 0.034 &  23.3 & 2.1e+11 & 2.0 & SF \\ 
WISEPC J170100.39+342042.9 &  255.25163 &   34.34524 & 0.101 & 6.27 & 0.039 &  23.1 & 2.1e+11 & 0.2 & SF \\ 
WISEPC J232514.19+151442.1 &  351.30911 &   15.24503 & 0.043 & 4.28 & 0.021 &  21.8 & 2.0e+11 & 0.3 & SF \\ 
WISEPC J165948.48+335944.1 &  254.95201 &   33.99559 & 0.085 & 5.96 & 0.036 &  20.4 & 1.9e+11 & 0.1 & SF \\ 
WISEPC J152138.79+305037.4 &  230.41161 &   30.84372 & 0.081 & 6.07 & 0.040 &  16.2 & 1.5e+11 & 1.0 & SF \\ 
WISEPC J104319.01+050818.0 &  160.82922 &    5.13833 & 0.068 & 5.70 & 0.041 &  15.4 & 1.5e+11 & 0.8 & SF \\ 
WISEPC J151941.93+312905.5 &  229.92471 &   31.48486 & 0.080 & 6.11 & 0.031 &  15.2 & 1.4e+11 & 1.1 & SF \\ 
WISEPC J221445.86-092300.8 &  333.69110 &   -9.38356 & 0.082 & 6.24 & 0.051 &  14.2 & 1.4e+11 & 0.4 & AGN \\ 
WISEPC J102200.74+382914.4 &  155.50308 &   38.48734 & 0.057 & 5.39 & 0.030 &  14.0 & 1.3e+11 & 0.5 & SF \\ 
WISEPC J130330.80-021400.0 &  195.87834 &   -2.23334 & 0.084 & 6.38 & 0.049 &  13.2 & 1.3e+11 & 0.1 & AGN \\ 
WISEPC J152021.87+485222.0 &  230.09113 &   48.87279 & 0.074 & 6.09 & 0.033 &  12.8 & 1.2e+11 & 1.5 & SF \\ 
WISEPC J170957.92+335507.3 &  257.49133 &   33.91869 & 0.084 & 6.42 & 0.042 &  12.5 & 1.2e+11 & 0.2 & SF \\ 
WISEPC J004336.31-092547.6 &   10.90130 &   -9.42988 & 0.050 & 5.29 & 0.030 &  11.3 & 1.1e+11 & 0.3 & SF \\ 
WISEPC J155857.04+273758.0 &  239.73769 &   27.63278 & 0.090 & 6.70 & 0.056 &  11.2 & 1.1e+11 & 1.7 & SF \\ 
WISEPC J170231.67+335135.6 &  255.63196 &   33.85988 & 0.087 & 6.61 & 0.051 &  11.2 & 1.1e+11 & 2.6 & SF \\ 
WISEPC J110018.01+100256.9 &  165.07504 &   10.04914 & 0.036 & 4.57 & 0.027 &  10.7 & 1.1e+11 & 0.8 & AGN \\ 
WISEPC J101346.84-005450.9 &  153.44518 &   -0.91414 & 0.042 & 4.96 & 0.028 &  10.6 & 1.0e+11 & 0.7 & AGN \\ 
WISEPC J163032.67+392303.2 &  247.63612 &   39.38421 & 0.030 & 4.21 & 0.024 &  10.5 & 1.0e+11 & 2.2 & AGN \\ 
WISEPC J122940.27+121743.6 &  187.41780 &   12.29546 & 0.087 & 6.72 & 0.155 &  10.2 & 1.0e+11 & 0.1 & SF \\ 
WISEPC J114623.86+552422.2 &  176.59941 &   55.40616 & 0.053 & 5.53 & 0.024 &  10.0 & 9.9e+10 & 1.7 & SF \\ 
WISEPC J005656.90-011242.4 &   14.23707 &   -1.21176 & 0.050 & 5.45 & 0.036 &   9.6 & 9.6e+10 & 0.3 & SF \\ 
WISEPC J162143.25+294332.6 &  245.43019 &   29.72572 & 0.098 & 7.09 & 0.070 &   9.5 & 9.4e+10 & 2.8 & SF \\ 
WISEPC J102126.42+381747.5 &  155.36008 &   38.29652 & 0.055 & 5.72 & 0.038 &   9.1 & 9.1e+10 & 1.2 & SF* \\ 
WISEPC J165903.80+334854.6 &  254.76585 &   33.81516 & 0.086 & 6.81 & 0.062 &   8.9 & 8.8e+10 & 0.2 & SF \\ 
WISEPC J171447.38+643541.1 &  258.69742 &   64.59475 & 0.080 & 6.67 & 0.050 &   8.7 & 8.7e+10 & 0.4 & SF \\ 
WISEPC J132632.23+002800.7 &  201.63428 &    0.46685 & 0.086 & 6.84 & 0.060 &   8.7 & 8.7e+10 & 0.3 & SF \\ 
WISEPC J075225.73+283040.0 &  118.10722 &   28.51112 & 0.062 & 6.08 & 0.044 &   8.5 & 8.5e+10 & 0.3 & AGN \\ 
WISEPC J103342.72+392926.9 &  158.42801 &   39.49082 & 0.068 & 6.32 & 0.049 &   8.3 & 8.4e+10 & 1.2 & SF \\ 
WISEPC J132944.92-014239.7 &  202.43716 &   -1.71104 & 0.089 & 6.98 & 0.077 &   8.2 & 8.2e+10 & 0.8 & SF \\ 
WISEPC J162345.89+410456.5 &  245.94122 &   41.08235 & 0.034 & 4.70 & 0.016 &   8.1 & 8.1e+10 & 2.1 & AGN \\ 
WISEPC J125830.21-015835.4 &  194.62585 &   -1.97650 & 0.080 & 6.75 & 0.073 &   8.1 & 8.1e+10 & 0.2 & SF* \\ 
WISEPC J115720.93+051506.0 &  179.33719 &    5.25168 & 0.081 & 6.80 & 0.064 &   8.0 & 8.0e+10 & 1.6 & SF \\ 
WISEPC J152003.58+304350.6 &  230.01491 &   30.73073 & 0.082 & 6.84 & 0.208 &   7.9 & 7.9e+10 & 1.7 & SF \\ 
WISEPC J155637.07+270043.1 &  239.15446 &   27.01196 & 0.091 & 7.11 & 0.079 &   7.7 & 7.7e+10 & 1.6 & SF \\ 
WISEPC J154351.50+363136.9 &  235.96458 &   36.52691 & 0.067 & 6.38 & 0.033 &   7.6 & 7.7e+10 & 1.4 & AGN \\ 
WISEPC J121754.98+040117.5 &  184.47910 &    4.02153 & 0.082 & 6.87 & 0.108 &   7.6 & 7.7e+10 & 2.1 & SF \\ 
WISEPC J102946.81+401913.7 &  157.44505 &   40.32048 & 0.067 & 6.39 & 0.047 &   7.6 & 7.6e+10 & 1.3 & AGN \\ 
WISEPC J162637.09+390739.3 &  246.65456 &   39.12757 & 0.035 & 4.89 & 0.016 &   7.5 & 7.6e+10 & 2.6 & SF \\ 
WISEPC J123011.93+120632.8 &  187.54971 &   12.10911 & 0.085 & 6.96 & 0.180 &   7.5 & 7.6e+10 & 0.1 & SF \\ 
WISEPC J112344.58+030018.9 &  170.93575 &    3.00525 & 0.051 & 5.75 & 0.036 &   7.5 & 7.5e+10 & 1.8 & SF \\ 
WISEPC J152255.30+305905.4 &  230.73042 &   30.98484 & 0.080 & 6.83 & 0.073 &   7.4 & 7.5e+10 & 1.4 & SF* \\ 
WISEPC J170102.30+340400.7 &  255.25958 &   34.06686 & 0.094 & 7.23 & 0.069 &   7.4 & 7.5e+10 & 0.2 & AGN \\ 
WISEPC J155842.84+270736.5 &  239.67848 &   27.12680 & 0.086 & 7.03 & 0.060 &   7.3 & 7.4e+10 & 1.6 & SF \\ 
WISEPC J115604.23+050150.7 &  179.01761 &    5.03075 & 0.079 & 6.81 & 0.075 &   7.3 & 7.4e+10 & 2.1 & SF* \\ 
WISEPC J152420.43+295719.7 &  231.08514 &   29.95546 & 0.076 & 6.72 & 0.049 &   7.3 & 7.3e+10 & 1.5 & SF \\ 
WISEPC J075213.93+292023.5 &  118.05803 &   29.33986 & 0.061 & 6.22 & 0.034 &   7.2 & 7.3e+10 & 0.4 & SF \\ 
WISEPC J133742.55+585209.8 &  204.42731 &   58.86938 & 0.074 & 6.68 & 0.078 &   7.2 & 7.2e+10 & 0.9 & SF \\ 
WISEPC J130344.14-030652.2 &  195.93390 &   -3.11451 & 0.081 & 6.91 & 0.081 &   7.1 & 7.2e+10 & 0.4 & SF \\ 
WISEPC J123018.51+113811.4 &  187.57713 &   11.63651 & 0.083 & 6.99 & 0.101 &   7.0 & 7.1e+10 & 2.5 & SF \\ 
WISEPC J155711.60+274455.2 &  239.29832 &   27.74868 & 0.088 & 7.12 & 0.073 &   7.0 & 7.1e+10 & 2.1 & SF \\ 
WISEPC J011242.10+010839.7 &   18.17543 &    1.14435 & 0.044 & 5.51 & 0.033 &   6.9 & 7.0e+10 & 0.5 & SF \\ 
WISEPC J161241.86+484805.8 &  243.17441 &   48.80160 & 0.059 & 6.16 & 0.034 &   6.9 & 7.0e+10 & 2.0 & SF \\ 
WISEPC J133940.58+590307.8 &  204.91910 &   59.05216 & 0.073 & 6.70 & 0.081 &   6.9 & 6.9e+10 & 0.8 & SF \\ 
WISEPC J165915.56+331019.5 &  254.81482 &   33.17208 & 0.087 & 7.13 & 0.072 &   6.8 & 6.9e+10 & 0.0 & SF \\ 
WISEPC J171424.22+633937.8 &  258.60092 &   63.66050 & 0.081 & 6.95 & 0.110 &   6.8 & 6.9e+10 & 0.1 & SF \\ 
WISEPC J135135.91+045100.4 &  207.89963 &    4.85010 & 0.076 & 6.78 & 0.069 &   6.8 & 6.9e+10 & 1.1 & SF \\ 
WISEPC J132633.84+000924.7 &  201.64101 &    0.15686 & 0.086 & 7.09 & 0.061 &   6.8 & 6.9e+10 & 0.5 & SF \\ 
WISEPC J111459.27+402142.0 &  168.74695 &   40.36167 & 0.073 & 6.70 & 0.060 &   6.8 & 6.9e+10 & 0.8 & SF \\ 
WISEPC J112016.54+540624.1 &  170.06892 &   54.10668 & 0.072 & 6.68 & 0.047 &   6.7 & 6.8e+10 & 1.4 & SF \\ 
WISEPC J170522.43+334145.0 &  256.34344 &   33.69582 & 0.090 & 7.23 & 0.077 &   6.6 & 6.7e+10 & 0.0 & SF \\ 
WISEPC J152043.23+304122.7 &  230.18015 &   30.68964 & 0.077 & 6.86 & 0.058 &   6.5 & 6.6e+10 & 0.9 & AGN \\ 
WISEPC J154256.63+351602.8 &  235.73595 &   35.26744 & 0.070 & 6.64 & 0.041 &   6.5 & 6.6e+10 & 1.7 & SF \\ 
WISEPC J170132.94+331901.1 &  255.38724 &   33.31697 & 0.088 & 7.18 & 0.069 &   6.4 & 6.6e+10 & 2.5 & AGN \\ 
WISEPC J152227.36+303446.0 &  230.61401 &   30.57945 & 0.078 & 6.90 & 0.051 &   6.4 & 6.6e+10 & 1.6 & SF \\ 
WISEPC J134227.74+585930.3 &  205.61560 &   58.99174 & 0.072 & 6.73 & 0.057 &   6.3 & 6.5e+10 & 0.6 & SF \\ 
WISEPC J235425.88-095923.4 &  358.60782 &   -9.98984 & 0.076 & 6.88 & 0.067 &   6.2 & 6.4e+10 & 0.9 & SF \\ 
WISEPC J151933.94+303057.6 &  229.89142 &   30.51601 & 0.078 & 6.93 & 0.071 &   6.2 & 6.3e+10 & 1.0 & SF \\ 
WISEPC J152418.93+295638.8 &  231.07889 &   29.94411 & 0.076 & 6.89 & 0.056 &   6.1 & 6.3e+10 & 1.2 & SF \\ 
WISEPC J111250.48+015616.5 &  168.21034 &    1.93791 & 0.076 & 6.91 & 0.077 &   6.1 & 6.2e+10 & 1.5 & SF \\ 
WISEPC J114951.78+560852.8 &  177.46574 &   56.14800 & 0.050 & 5.93 & 0.040 &   6.0 & 6.2e+10 & 1.5 & SF \\ 
WISEPC J162549.24+402042.8 &  246.45515 &   40.34522 & 0.029 & 4.70 & 0.028 &   5.9 & 6.0e+10 & 2.6 & SF \\ 
WISEPC J111925.05+545849.5 &  169.85439 &   54.98043 & 0.071 & 6.78 & 0.056 &   5.9 & 6.0e+10 & 1.6 & SF \\ 
WISEPC J122747.50+120322.3 &  186.94794 &   12.05620 & 0.088 & 7.28 & 0.073 &   5.9 & 6.0e+10 & 0.1 & SF \\ 
WISEPC J111459.91+024551.0 &  168.74963 &    2.76418 & 0.077 & 6.95 & 0.056 &   5.9 & 6.0e+10 & 1.3 & SF \\ 
WISEPC J162253.90+402947.0 &  245.72458 &   40.49639 & 0.031 & 4.85 & 0.026 &   5.9 & 6.0e+10 & 2.6 & SF \\ 
WISEPC J155849.70+272639.2 &  239.70708 &   27.44421 & 0.085 & 7.20 & 0.076 &   5.9 & 6.0e+10 & 1.8 & SF \\ 
WISEPC J152308.37+304847.1 &  230.78488 &   30.81309 & 0.072 & 6.82 & 0.064 &   5.9 & 6.0e+10 & 1.4 & SF \\ 
WISEPC J102815.22+400800.0 &  157.06342 &   40.13334 & 0.067 & 6.65 & 0.064 &   5.8 & 5.9e+10 & 0.8 & SF \\ 
WISEPC J141309.46+442850.4 &  213.28941 &   44.48066 & 0.090 & 7.37 & 0.082 &   5.7 & 5.9e+10 & 1.1 & SF \\ 
WISEPC J011703.58+000027.4 &   19.26492 &    0.00762 & 0.046 & 5.76 & 0.029 &   5.7 & 5.9e+10 & 0.5 & AGN \\ 
WISEPC J111341.36+412319.9 &  168.42235 &   41.38885 & 0.073 & 6.88 & 0.062 &   5.7 & 5.9e+10 & 0.8 & SF \\ 
WISEPC J102134.50+410605.2 &  155.39377 &   41.10143 & 0.092 & 7.42 & 0.103 &   5.7 & 5.8e+10 & 0.5 & SF \\ 
WISEPC J132459.25+110431.7 &  201.24689 &   11.07547 & 0.084 & 7.21 & 0.142 &   5.7 & 5.8e+10 & 0.5 & SF \\ 
WISEPC J111512.08+542741.4 &  168.80035 &   54.46151 & 0.066 & 6.64 & 0.057 &   5.7 & 5.8e+10 & 1.5 & SF \\ 
WISEPC J133450.77+593713.3 &  203.71155 &   59.62035 & 0.076 & 6.99 & 0.075 &   5.6 & 5.8e+10 & 0.9 & SF \\ 
WISEPC J132054.04+113912.6 &  200.22517 &   11.65351 & 0.095 & 7.51 & 0.147 &   5.6 & 5.8e+10 & 0.6 & SF \\ 
WISEPC J103130.85+401450.9 &  157.87852 &   40.24746 & 0.066 & 6.66 & 0.058 &   5.6 & 5.7e+10 & 1.3 & SF \\ 
WISEPC J155844.97+270812.4 &  239.68739 &   27.13678 & 0.095 & 7.53 & 0.095 &   5.5 & 5.7e+10 & 1.8 & SF \\ 
WISEPC J104115.63+045313.8 &  160.31512 &    4.88717 & 0.068 & 6.76 & 0.063 &   5.5 & 5.6e+10 & 1.3 & AGN \\ 
WISEPC J162036.15+335643.1 &  245.15063 &   33.94529 & 0.031 & 4.97 & 0.026 &   5.4 & 5.6e+10 & 2.1 & SF \\ 
WISEPC J130306.96-031846.9 &  195.77899 &   -3.31304 & 0.086 & 7.30 & 0.112 &   5.4 & 5.6e+10 & 0.2 & SF \\ 
WISEPC J115809.58+555140.0 &  179.53992 &   55.86110 & 0.063 & 6.57 & 0.084 &   5.4 & 5.5e+10 & 1.8 & SF \\ 
WISEPC J171248.41+634258.0 &  258.20169 &   63.71612 & 0.075 & 7.01 & 0.067 &   5.4 & 5.5e+10 & 0.3 & SF \\ 
WISEPC J115938.06+561702.5 &  179.90857 &   56.28402 & 0.070 & 6.85 & 0.080 &   5.3 & 5.4e+10 & 2.2 & SF* \\ 
WISEPC J121734.75+035019.8 &  184.39478 &    3.83884 & 0.073 & 6.96 & 0.110 &   5.3 & 5.4e+10 & 1.9 & SF \\ 
WISEPC J155745.93+274043.6 &  239.44136 &   27.67878 & 0.090 & 7.46 & 0.096 &   5.2 & 5.4e+10 & 1.7 & SF \\ 
WISEPC J101051.19-003842.5 &  152.71327 &   -0.64514 & 0.043 & 5.75 & 0.076 &   5.2 & 5.4e+10 & 0.8 & SF* \\ 
WISEPC J130510.25-023516.0 &  196.29272 &   -2.58778 & 0.088 & 7.41 & 0.133 &   5.2 & 5.4e+10 & 0.4 & SF \\ 
WISEPC J162858.00+391909.3 &  247.24168 &   39.31924 & 0.034 & 5.18 & 0.027 &   5.2 & 5.3e+10 & 2.4 & SF \\ 
WISEPC J151848.57+305449.2 &  229.70239 &   30.91367 & 0.078 & 7.13 & 0.064 &   5.2 & 5.3e+10 & 1.2 & SF \\ 
WISEPC J123001.03+113940.5 &  187.50427 &   11.66124 & 0.082 & 7.24 & 0.124 &   5.2 & 5.3e+10 & 2.2 & SF \\ 
WISEPC J133105.70+584939.1 &  202.77376 &   58.82752 & 0.073 & 6.97 & 0.065 &   5.1 & 5.3e+10 & 0.8 & SF \\ 
WISEPC J152428.76+310750.4 &  231.11983 &   31.13067 & 0.078 & 7.14 & 0.094 &   5.1 & 5.2e+10 & 1.1 & AGN \\ 
WISEPC J103954.63+051546.3 &  159.97762 &    5.26285 & 0.064 & 6.69 & 0.069 &   5.1 & 5.2e+10 & 1.7 & SF \\ 
WISEPC J132422.33+105047.7 &  201.09305 &   10.84658 & 0.093 & 7.59 & 0.175 &   5.0 & 5.2e+10 & 0.7 & SF \\ 
WISEPC J082926.09+302528.8 &  127.35872 &   30.42466 & 0.051 & 6.17 & 0.047 &   5.0 & 5.2e+10 & 0.7 & SF \\ 
WISEPC J111503.18+542856.5 &  168.76323 &   54.48237 & 0.067 & 6.80 & 0.076 &   5.0 & 5.2e+10 & 1.5 & SF \\ 
WISEPC J155957.42+272745.9 &  239.98924 &   27.46274 & 0.095 & 7.63 & 0.124 &   5.0 & 5.1e+10 & 1.8 & AGN \\ 
WISEPC J171351.15+640012.7 &  258.46313 &   64.00354 & 0.088 & 7.45 & 0.179 &   5.0 & 5.1e+10 & 0.2 & SF \\ 
WISEPC J155930.98+273257.4 &  239.87907 &   27.54927 & 0.091 & 7.53 & 0.132 &   4.9 & 5.1e+10 & 1.9 & SF \\ 
WISEPC J235345.66-103353.6 &  358.44025 &  -10.56490 & 0.078 & 7.19 & 0.083 &   4.9 & 5.1e+10 & 0.7 & SF \\ 
WISEPC J115610.78+045943.3 &  179.04491 &    4.99536 & 0.079 & 7.24 & 0.104 &   4.8 & 5.0e+10 & 1.6 & AGN \\ 
WISEPC J122915.88+113858.6 &  187.31615 &   11.64962 & 0.088 & 7.49 & 0.160 &   4.8 & 4.9e+10 & 2.5 & SF \\ 
WISEPC J235514.75-102946.9 &  358.81146 &  -10.49636 & 0.082 & 7.33 & 0.105 &   4.7 & 4.9e+10 & 0.6 & SF \\ 
WISEPC J133111.02-014338.6 &  202.79590 &   -1.72740 & 0.084 & 7.39 & 0.096 &   4.7 & 4.9e+10 & 0.7 & SF \\ 
WISEPC J235208.99-095434.4 &  358.03748 &   -9.90957 & 0.077 & 7.19 & 0.082 &   4.7 & 4.9e+10 & 0.6 & SF* \\ 
WISEPC J134033.01+015016.0 &  205.13754 &    1.83778 & 0.078 & 7.24 & 0.094 &   4.7 & 4.9e+10 & 0.5 & SF \\ 
WISEPC J130335.10-025417.0 &  195.89626 &   -2.90471 & 0.083 & 7.38 & 0.110 &   4.7 & 4.8e+10 & 0.2 & SF \\ 
WISEPC J130329.04-025337.2 &  195.87102 &   -2.89366 & 0.079 & 7.27 & 0.092 &   4.7 & 4.8e+10 & 0.3 & SF \\ 
WISEPC J122902.48+114340.3 &  187.26031 &   11.72786 & 0.086 & 7.49 & 0.243 &   4.6 & 4.8e+10 & 2.3 & SF \\ 
WISEPC J135143.98+045833.5 &  207.93326 &    4.97596 & 0.077 & 7.20 & 0.089 &   4.6 & 4.8e+10 & 0.5 & SF \\ 
WISEPC J170436.61+334454.2 &  256.15256 &   33.74839 & 0.090 & 7.60 & 0.099 &   4.6 & 4.8e+10 & 3.0 & SF \\ 
WISEPC J121734.57+033450.9 &  184.39404 &    3.58081 & 0.074 & 7.13 & 0.089 &   4.6 & 4.8e+10 & 1.8 & SF \\ 
WISEPC J123102.64+113929.5 &  187.76102 &   11.65819 & 0.087 & 7.52 & 0.126 &   4.6 & 4.7e+10 & 2.8 & SF \\

\label{table:demi_lirgs}
\end{longtable}
\end{center}

\end{document}